\newif\iftr
\newif\ifcnf

\trtrue
\cnffalse

\iftr
  \documentclass[sigplan,screen,nonacm,10pt]{acmart}
  \startPage{1}
  \setcopyright{none}
  \acmConference[]{}{}{}
  \acmISBN{}
  \acmDOI{}
\else
  \documentclass[sigplan,screen,10pt]{acmart}
  \acmConference[PPoPP'2021]{}{Feb 27-Mar 3, 2021.}{Seoul, South Korea}
  \acmYear{2021}
  \acmISBN{} 
  \acmDOI{} 
  \setcopyright{none}
\fi

\usepackage{booktabs}
\usepackage{algorithmic}
\usepackage{adjustbox}
\usepackage{graphicx}
\usepackage{subfig}
\usepackage{textcomp}
\usepackage{wrapfig}
\usepackage{import}
\usepackage{xcolor}
\usepackage{float}
\usepackage{siunitx}
\usepackage[switch]{lineno}
\usepackage{titlesec}
\def\BibTeX{{\rm B\kern-.05em{\sc i\kern-.025em b}\kern-.08em
    T\kern-.1667em\lower.7ex\hbox{E}\kern-.125emX}}

\newcommand{\caseicon}[1]{
  \texorpdfstring{
    \protect\raisebox{
      -.3\baselineskip
    }{
      \protect\includegraphics[scale=0.5,trim=0 0 0 0, clip]{#1}\hspace{-0.2cm}
    }
  }{}
}

\usepackage{xargs}

\usepackage{booktabs}

\usepackage{listings}
\usepackage{multirow}

\newcommand{\code}[1]{\texttt{#1}}
\usepackage{xspace}
\newcommand{\toolname}{Perf-Taint\xspace}
\newfloat{codeblock}{H}{myc}
\definecolor{darkblue}{rgb}{0,0,.6}
\definecolor{darkred}{rgb}{.6,0,0}
\definecolor{darkgreen}{rgb}{0,.6,0}
\definecolor{red}{rgb}{.98,0,0}
\definecolor{gray}{rgb}{.6,.6,.6}
\newtheorem{theorem}{Theorem}
\newtheorem{claim}{Claim}

\begin{document}

\title[Extracting Clean Performance Models from Tainted Programs]{Extracting Clean Performance Models\\ from Tainted Programs}

\author{Marcin Copik}
\email{marcin.copik@inf.ethz.ch}
\affiliation{%
  \institution{Department of Computer Science, ETH Zurich}
}

\author{Alexandru Calotoiu}
\email{alexandru.calotoiu@inf.ethz.ch}
\affiliation{%
  \institution{Department of Computer Science, ETH Zurich}
}

\author{Tobias Grosser}
\email{tobias.grosser@ed.ac.uk}
\affiliation{%
  \institution{School of Informatics, University of Edinburgh}
}

\author{Nicolas Wicki}
\email{nwicki@ethz.ch}
\affiliation{%
  \institution{Department of Computer Science, ETH Zurich}
}

\author{Felix Wolf}
\email{wolf@cs.tu-darmstadt.de}
\affiliation{%
  \institution{Department of Computer Science, Technical University of Darmstadt}
}

\author{Torsten Hoefler}
\email{htor@inf.ethz.ch}
\affiliation{%
  \institution{Department of Computer Science, ETH Zurich}
}

\begin{abstract}

  Performance models are well-known instruments to understand the
  scaling behavior of parallel applications. They express how
  performance changes as key execution parameters, such as the number
  of processes or the size of the input problem, vary. Besides
  reasoning about program behavior, such models can also be
  automatically derived from performance data. This is called
  empirical performance modeling. While this sounds simple at the
  first glance, this approach faces several serious interrelated
  challenges, including expensive performance measurements,
  inaccuracies inflicted by noisy benchmark data, and overall
  complex experiment design, starting with the selection of the
  right parameters. The more parameters one considers, the more
  experiments are needed and the stronger the impact of noise. In this
  paper, we show how taint analysis, a technique borrowed from the
  domain of computer security, can substantially improve the modeling
  process, lowering its cost, improving model quality, and help
  validate performance models and experimental setups.

\end{abstract}

\maketitle

\section{Introduction}
The increasing complexity of both hardware and scientific problems
creates new challenges for developers of high-performance
applications. The design process of a massively parallel program that
can scale on modern architectures requires a deep understanding of
computational kernels and communication patterns. Performance modeling
has become a standard technique to solve problems such as locating
scalability bottlenecks~\cite{automated-modeling,Siegmund:2015:PMH:2786805.2786845,Goldsmith:2007:MEC:1287624.1287681}, estimating the
execution time when the input size or the core count changes~\cite{hoisie-palm-2012}, or
predicting  application performance on a new
architecture~\cite{modeling-codesign,lo2014roofline}.
	

The main goal of performance modeling is to express the performance of an
application as a function of one or more execution
parameters~\cite{1592815, Hoefler:2011:PMS:2063348.2063356}. Purely
analytical performance modeling involves an expert who analyzes the
source code and understands the underlying
algorithms~\cite{Hoefler:2011:PMS:2063348.2063356}. While very
effective once the models have been created, the required person-hours
and experience restrict its usability in practice.
Empirical performance modeling, by contrast, generates similar
performance models automatically by analyzing measurements taken from
running an instrumented version of the application in different
configurations. It follows three major steps: identifying parameters,
designing an experiment to measure the influence of parameter changes
on the application behavior, and learning the model that best fits the
data. While generating models from existing data is automatic and
resource efficient, running the experiments may require careful
planning and extensive computational effort.
\begin{figure}
  \includegraphics[width=\linewidth]{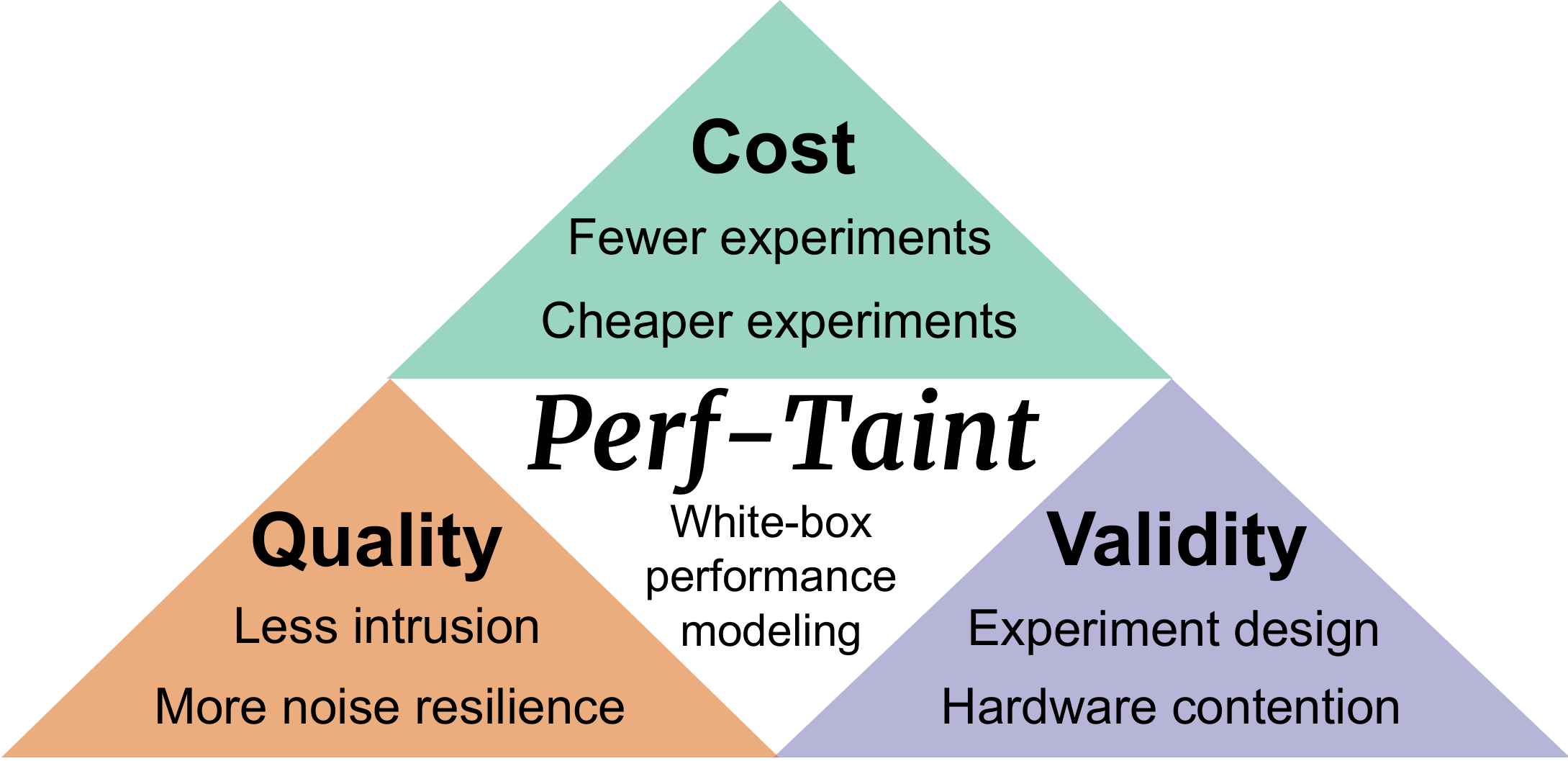}
  \caption{
    The basic concept of \toolname. A tainted run of the
    program provides information that improves the empirical
    performance modeling process along the dimensions cost, quality,
    and validity of the resulting models.
  }
  \label{fig:system-overview}
\end{figure} 
In general, empirical performance modeling involves two important
decisions: (1) choosing parameters that will affect application
performance and (2) designing a set of experiments capable of
accurately measuring their influence, while not
exhausting the available computational budget. 

Modern scientific
applications use dozens of parameters that describe numerical
properties, data size,
or the degree of
parallelism, making their selection extremely challenging. Without
detailed insight into the application behavior, the user has to
consider all possible combinations of the chosen parameters. The larger
the number of parameters, the bigger is the number of experiments and the
impact of noise on the quality of the resulting
models~\cite{ritter_ea:2020:ipdps}.  Some performance effects are not
measurable for the entire range of parameter values available in
the experiment design, potentially invalidating some measurements.
Another major difficulty arises from the black-box nature of empirical
modeling. Without insight into function behavior beyond empirical
data, the modeler cannot distinguish between actual runtime change
because of parameter influence and the effects of noise on
the measurements. This leads to overfitting, estimating false
dependencies, and generating incorrect models for constant functions
with negligible execution time.

In this paper, we show how taint analysis~\cite{clause2007dytan},
a technique borrowed from the field of computer
security, which reliably relates marked input values with the program
parts they potentially affect, can provide this additional context,
leading to the concept of {\em tainted performance
  modeling}. Performance tainting provides us with accurate
performance parameter information and enables the design of a novel
loop-based complexity analyzer. We integrate our
complexity analyzer with Extra-P~\cite{automated-modeling}, an
empirical performance-modeling tool, and derive a new hybrid
performance modeling framework called {\em \toolname}, whose underlying
concept is illustrated in Figure~\ref{fig:system-overview}. We make
the following specific contributions:

\begin{itemize}

\item The application of taint analysis, which has originally been
  devised to track the flow of protected data through a program, to
  a new problem: the improvement of empirical performance models of HPC
  applications.

\item The elaborate concept of tainted performance modeling, that can (1)
  reduce the cost of empirical performance models, (2) improve their
  quality, and (3) help validate them and the experimental setups used
  for their generation. An example of such a validation is the
  detection of contention as the source of measurements that contradict
  expected computational volumes.

\item An open-source, LLVM-based tainted performance
  modeling tool, ready to use on HPC programs to provide the insights of a
  virtual performance expert\footnote{The code is available on GitHub: \code{spcl/perf-taint}}.
 
\end{itemize}

After reviewing related work in \autoref{sec:rl}, and the taint
analysis in \autoref{sec:dynamic_taint_analysis}, we describe the
theoretical and practical aspects of our method
in \autoref{sec:performance-modeling}
and \autoref{sec:perf_taint}, respectively. We then demonstrate
the benefits it provides in \autoref{sec:case-studies} and present our
conclusion in \autoref{sec:conclusion}.

\section{Related Work}

\label{sec:rl}

The broad spectrum of existing methods and tools to support the creation of performance models documents their importance for understanding the performance influence of algorithms~\cite{Zaparanuks:2012:AP:2345156.2254074}, the hardware~\cite{lo2014roofline,Marin:2004:CPP:1012888.1005691}, and the operating system~\cite{DBLP:journals/corr/HammerHEW15,Zaparanuks:2012:AP:2345156.2254074,Meswani:2013:MPP:2493921.2493922}. They are often used to extrapolate performance outside the known range of a single parameter~\cite{Zhai:2010:PPP:1837853.1693493,Vuduc:2004:SME:1080647.1080688,7161600} or even multiple parameters~\cite{Siegmund:2015:PMH:2786805.2786845,Goldsmith:2007:MEC:1287624.1287681}, sometimes exploiting the properties of certain classes of algorithms such as stencil computations~\cite{wu_ea:2012}. Some require the prior annotation of the code with performance expressions~\cite{hoisie-palm-2012,vetter-aspen-2012}. Machine learning methods have also been successively used for performance modeling~\cite{Ipek:2005:APP:2138773.2138800,Lee:2007:MIL:1229428.1229479}.

There have been several attempts to enable performance modeling through static
analysis of source code~\cite{1303333,8048922,Lee:2015:CFA:2751205.2751220,8385238}.
Thanks to the dynamic nature of taint analysis, our method is not affected by fundamental limitations of static methods. The dependence of performance modeling tools on an entirely static dataflow analysis or a perfect loop modeling might prevent them from scaling to large scientific applications.
A hybrid performance modeling tool was presented for online modeling by Bhattacharyya et al.~\cite{7429329}.

A different aspect of performance modeling, dataflow analysis, is also well studied in high-performance computing:
DfAnalyzer performs dynamic dataflow analysis of Spark high-performance applications~\cite{vldb}.
Parallel control-flow graphs of MPI programs have been constructed with dataflow
analysis~\cite{Aananthakrishnan:2013:HAD:2464996.2467286}.
Value influence analysis, a variant of taint analysis, has been used in message-passing
applications~\cite{Roth:2014:VIA:2597652.2597666}.

\section{Dynamic Taint Analysis}
\label{sec:dynamic_taint_analysis}

When trying to construct the performance model of a full application,
one would hope that there is a way to automatically determine its
runtime complexity (or complexity w.r.t. other metrics) by analyzing
the program code with a sufficiently smart compiler. While previous
research showed first results~\cite{hoefler-noise-sim} towards this
direction, these solutions are inherently limited to special cases or
approximations and are hard to scale to non-trivial programs
(\autoref{sec:static_modeling}), as even the simple problem of
identifying which input parameter affects an arbitrary program
variable is inherently difficult.

Dynamic taint analysis (\autoref{sec:taint_analysis}) has been
successfully used in the context of computer security to reliably
analyze data relations across complex programs.
We introduce the major concepts and techniques and discuss how taint analysis
provides precise knowledge as to how input parameters affect
variables of the program. This gives us the instruments to 
introduce white-box performance modeling of non-recursive HPC programs 
(\autoref{sec:performance-modeling}).

\subsection{Static performance modeling is hard}
\label{sec:static_modeling}

Modeling performance statically is difficult from both a theoretical and
a practical viewpoint. There are strict theoretical limitations of
how accurate a static analysis can be or if an analysis can be computed at
all.
On the practical side, various levels of
abstractions or indirections make program code easy to maintain, but make static analysis even less likely to succeed.
We now outline several theoretical and practical considerations that
make static performance modeling hard.

\paragraph{Theoretic limitations} 

There are strong theoretic reasons why static program analysis techniques are
often unable to provide precise answers even for seemingly simple analyses.
The most well known is the halting problem~\cite{church1936unsolvable} or its
generalization, Rice's theorem~\cite{rice1953classes}, which we here rephrase
in the terms of program analysis: any non-trivial semantic property of a
program cannot be computed. A property such as ``does a program contain a
certain instruction sequence'' is syntactic and might be decidable, but
``does a program return 0'' is a semantic property and is undecidable for
arbitrary programs.  As a result, the question ``can the value
stored in a given memory location affect an instruction that is run when
executing a given program'' is a semantic property. Hence, determining whether a
configuration parameter affects certain parts of a program is
undecidable. 

\paragraph{Practical considerations}

While Rice's theorem shows that the proof of semantic properties
to be impossible for all programs, there might still exist a sufficiently
large set of programs where this is realistic. We now argue that
even an approximate analysis
for only a subset of programs is difficult in practice. The main culprits are abstraction
overhead, complex abstract data types, and runtime configurability.
\ifcnf
\else
Abstraction is important to ensure the maintainability of large
software projects. To that end, class hierarchies, virtual dispatching, and
many very fine-granular functions are commonly used.
As an example we show a simple matrix initialization in C99 and C++ (even avoiding the use of modern C++
constructs such as lambdas, range loops, etc.):
%
\begin{lstlisting}[language=C++]
void init_c99(int Rows, int Columns,
              float A[Rows][Columns], float val) {
  for (int i = 0; i < Rows; i++)
    for (int j = 0; j < Columns; j++)                                          
      A[i][j] = val;
}
    
class matrix {
  int Rows, Columns;
  float *A;
public:                                                                        
  float &at(unsigned i, unsigned j) {
    return A[i * Columns + j];
  }
  unsigned getColumns() { return Columns; }                                    
  unsigned getRows() { return Rows; }                                          
};
void init_cpp(matrix &A, float val) {                                    
  for (int i = 0; i < A.getRows(); i++)                                        
    for (int j = 0; j < A.getColumns(); j++)                                   
      A.at(i,j) = val;                                                         
}                                                                              
\end{lstlisting}
%
While the initialization routine looks very similar in both cases, deriving the number of loop
iterations is significantly more complicated for the C++ function. In the C99 variant, the
number of rows in the matrix $A$ is directly available as a raw integer
variable. In contrast, the C++ variant of this code encapsulates this information in
the \code{matrix} data structure. Compilers are certainly good at reducing abstraction
overheads and especially C++ is known as a language to make this easy.
However, compilers such as LLVM or gcc do not reliably eliminate the
abstraction overhead even in this simple example. There are two main problems
that make eliminating this
overhead difficult: (1) the matrix itself is passed as a pointer,
and the size of the matrix is stored in
memory as member of the struct describing an instance of the class \code{matrix}.
Simplifying code with double pointer indirection requires a very good understanding of pointer
aliasing; (2) the type of the matrix is part of the calling convention, hence
the compiler is not allowed to change it if it appears in a library interface.

Pointer aliasing in general is a hard problem, where the most advanced
inter-procedural techniques~\cite{steensgaard1996points} are still very
imprecise~\cite{shapiro1997effects} on large code bases. In practice (e.g.,
LLVM), only very limited intra-procedural pointer-alias analyses are run, as
the minor increase in precision does not justify the large compile-time cost to
run them. Similarly, today's compilers only carefully inline
function calls as otherwise code size (as well as compile time and runtime)
grows uncontrollable. As a result, almost any analysis becomes inter-procedural
and in this setting virtual function calls (or exceptions) make it even more
challenging to understand statically how data flows in a program.

\fi
Another source of
over-approximation
is the hard-to-predict control-flow found
in languages gaining popularity in scientific computing.
The problem arises from virtual dispatch in statically typed languages (C++, Julia) and 
from duck typing (Python).

Message-passing and multithreading adds
to the code non-determinism that can lead to a combinatorial explosion
of states. A common choice for the analysis of MPI programs is
symbolic execution~\cite{7027430}, but it suffers from the exponential
number of paths it has to analyze, limiting its
scalability~\cite{10.1145/3183440.3190336}.

\subsection{Dynamic taint analysis}
\label{sec:taint_analysis}

Dynamic taint analysis is a runtime analysis that \emph{marks and tracks the
movement of certain data elements and computed results depending on them through the
execution of a program}. Taint analysis can compute semantic analyses
while a program is executed, overcoming the limitations that prevent static
(compile-time) analysis from providing precise results, at the price
of narrowing the insights to a specific run and its input
configuration.
For many analyses -- especially if the results anyhow depend on input data --
this is often exactly what is desired.
We introduce a general taint-analysis framework
that can be used to instantiate problem-specific taint analyses.
We define three major components described by Clause et al.~\cite{clause2007dytan} -- (a) taint sources,
(b) propagation policy, (c) taint sinks, which we discuss below with code examples.


\paragraph{Taint sources} Taint sources are all components of a
program that can represent some kind of program data. Typical taint
sources are memory locations, variable names, or function return
values, but almost any part of a program can
be a taint source, including I/O interfaces, system calls, network devices,
etc. Marking taint sources requires the specification of data to be
tainted and taint labels used to mark it.
\iftr
\begin{minipage}{\linewidth}
\fi
\begin{lstlisting}[basicstyle=\scriptsize]
// Program input: taint with label "size"
scanf("%d", &size); 
// Manual taint source: taint with label "p"
write_label(&p, sizeof(p), "p");
// Third-party library output: taint with label "ranks"
MPI_Comm_size(MPI_COMM_WORLD, &ranks);
\end{lstlisting}
\iftr
\end{minipage}
\fi

\paragraph{Taint propagation policy} The taint propagation policy defines how
taint labels are moved through a program. We specify it by (1) defining a mapping function,
and by (2) defining the affected data. The \emph{mapping function} defines how two (or more) taint labels are joined.
In the most trivial case, two sets of labels are joined by taking the
union of the sets.

The \emph{affected data} defines all data to which taint labels
are propagated through data-flow and control-flow.
%
Data-flow based propagation passes taint labels from
inputs of operations to their outputs, including program instructions
and propagation from function arguments to its return value.
Control-flow based propagation captures the propagation of taint
labels through control dependencies~\cite{DBLP:conf/ndss/KangMPS11,10.1145/1273463.1273490}.
\begin{lstlisting}[basicstyle=\scriptsize]
int foo(int a,int b,int c) {//Input labels: "a","b","c"
  int d = 2 * a; //Dataflow tainting with label "a"
  if(b) d++; else d--; //Explicit tainting, label "b"
  if(c) d = pow(d, 2); //Implicit tainting, label "c"
  return d; //Taint labels of return: "a", "b", "c"
}
\end{lstlisting}
In the example above, the variable \code{a} taints the
return value of the function through a data-flow propagation.
Variable \code{b} taints \code{d} through a control-flow condition
which controls the execution of code that changes the
return value (explicit dependence). An implicit dependence occurs for \code{c} since the value depends on
it even if the second branch is not taken.
\paragraph{Taint sinks} Taint sinks are program code locations with an
associated variable or memory location that may observe tainted program
data.
Sinks are used to
determine which tainted values affect a given behavior.
Each taint sink is defined by (1) a program code
location, (2) the variable or memory location to check, and (3) a checking
method that is invoked whenever the taint sink is executed.
\iftr
The checking method is supplied with program variable and associated taint labels on invocation.
Whenever a taint
sink's checking method is invoked, it is supplied with the variable or memory
location to check together with the set of associated labels.
\begin{lstlisting}[basicstyle=\scriptsize]
if(result > threshold) {
  handle_edge_case();
  sink({&result, &threshold}, edge_case_handler);
}
\end{lstlisting}
\fi

\section{Tainted Performance Modeling}
\label{sec:performance-modeling}

Building on the concepts presented in~\autoref{sec:taint_analysis},
we introduce a \emph{taint analysis for performance modeling}
(\autoref{sec:taint_analysis_for_perf_modeling}), where the influence of input parameters on program variables is used to model
the number of loop iterations. We show how this knowledge allows
us to place a bound on the \emph{volume of computation} (\autoref{sec:iteration_volume}), defined as the number of
operations executed in a non-recursive program run.
An empirical black-box performance modeler uses the information of compute volumes to limit the space of
potential complexity functions it considers
(\autoref{sec:hybrid-model}).

\subsection{Loop count parameter identification}
\label{sec:taint_analysis_for_perf_modeling}

We solve the core data-flow problem behind performance modeling by
formulating a suitable taint analysis. We assume that all target
metrics, such as the program runtime or the number of operations,
only vary with the iteration number of loop constructs in the code.
This assumption is intuitive because the source code is typically not changed
(i.e., in complexity or size) when changing program input parameters. 
Our analysis computes how potential
input parameters affect the iteration counts of all \textit{natural}
loops~\cite{Aho:1986:CPT:6448} in a program.\footnote{Our analysis does not explicitly
consider irreducible loops where control is transferred through multiple
paths into the loop (no single loop header), as irreducible loops can easily
be transformed into natural loops~\cite{unger2002handling}.}
While the analysis does not support recursive
  functions, it warns of over-approximation when recursion is detected. Nevertheless,
the core focus of performance modeling are HPC applications where the vast majority
of computations are iterative anyway.

\paragraph{Sources} The sources of our loop taint analysis are all potentially
performance-relevant parameters of a program. Performance relevant parameters
are all memory locations marked explicitly by the performance engineer with a
parameter label. Parameters are typically read from the command line, but might
also be provided through other means (e.g., a configuration file),
as long as their value is eventually stored in a variable that the user has
marked as a parameter.

\paragraph{Propagation policy} To reliably produce accurate results, our analysis
requires the propagation of taint across data flow and control flow.
Because we need to know only the presence of a specific taint
label in a performance-relevant variable, we choose the set union as mapping function.
Each label will contain the set of input parameters that in some way
affected the value marked by the taint label.

\paragraph{Sinks} The sinks of our taint analysis are all loop exit conditions.
For a given loop, the number of times it iterates depends only on
loop exit branch conditions. Any further
indirect dependencies will eventually taint these branch conditions through
our taint analysis.

We summarize the concepts in an example. With automatic taint propagation,
the dependency on input parameters is propagated through function calls and memory
operations to the sink, where it is used by our loop-count analysis.
\iftr
\begin{minipage}{\linewidth}
\fi
\begin{lstlisting}[basicstyle=\scriptsize]
struct params = parse_args();
write_label(&params.size,"size", &params.step,"step");
iterate(pow(params.size, 2), optimize_step(params));
void iterate(int size, int step) {
 for(int i = 0; i < size; i += step) {
  compute();  sink({&i, &size}, register_loop);
 }
}
\end{lstlisting}
\iftr
\end{minipage}
\fi
Assuming a loop $L$ depends on taint labels $p_1, \dots, p_n$, the number of
loop iterations of $L$, $count(L)$, must then be a function $g(p_1, \dots, p_n)$.
While the parameters this function potentially depends on are clearly defined,
no further information about function $g$ can be derived through
the taint analysis itself. Even if the taint label just contains a single taint mark
$p$, it does not imply a number of loop iterations linear in $p$. $count(L)$ could also be $log(p)$, $p^2$, or any other function $g(p)$. As a result, we can state the
following claim:

\begin{claim} Given an application with a set of input variables V and a
set of $n$ correctly marked potential performance variables $P = (p_1, \dots,
p_n) \subseteq V$, we derive for a given loop $L$ a class of symbolic functions
$g_i(p_1, \dots, p_n)$ which only depend on parameters in $P$. If all program
parameters that impact the iteration count of $L$ have been marked, a
performance taint analysis with full data and control flow propagation computes
with $count(L) = p_i$ a class of functions which contain the function that
exactly describes the number of loop iterations.
\end{claim}

\subsection{Iteration volume of a loop nest}
\label{sec:iteration_volume}

We now derive the iteration volume of a loop nest, that is, the accumulated
number of times the body of a loop nest is executed. 
We define the iteration volume recursively. The base case of
our recursion is a loop nest with just a single loop $L$. In this case, the
volume of computation $vol(L)$ is $count(L) = g(\vec{p})$, the result of our
loop iteration count parameter identification. We now define the iteration volume of
larger loop nests by combining existing loop nests.

\paragraph{Sequencing two loops} Executing two child loops nests ($LN_{c1}$ and
$LN_{c2}$) in sequence forms a larger loop nest $LN$. The iteration volume of
$LN$ is over-approximated as the sum of the compute volumes of the child loop
nests, $vol(LN) = vol(LN_{c1}) + vol(LN_{c2})$.

\begin{lstlisting}[basicstyle=\scriptsize]
LN:  {
        for(int i = 0; i < count(LN_c1); i += 1) { ... }
        for(int i = 0; i < count(LN_c2); i += 1) { ... }
     }
\end{lstlisting}

\paragraph{Nesting of a loop and a loop nest} Executing a child loop nest
($LN_c$) inside a loop $L$ forms a larger loop nest $LN$. The iteration volume
of $LN$ is over-approximated by multiplying the iteration count of the outer
loop with the iteration count of the loop nest, $vol(LN) = g(\vec{p}) \cdot
vol(LN_c)$.

\begin{lstlisting}[basicstyle=\scriptsize]
LN:  {
L_1:     for(int i = 0; i < count(L_0); i += 1)
           for(int j = 0; j < count(LN_child); j += 1)
             ...
     }
\end{lstlisting}

The power of these simple composition rules can be summarized in the
following claim about asymptotic performance with respect to performance-critical variables.
\begin{claim} Given an application with a set of input variables V, a
set of $n$ correctly marked potential performance variables $P = (p_1, \dots,
p_n) \subseteq V$, and a loop nest built from natural loops $L_i$ without
irreducible control flow or recursion, we derive for the loop nest a class of
symbolic functions $g_i(p_1, \dots, p_n)$ which only depend on the parameters
in $P$. If all variables that impact the loop iteration count have been
marked, we derive an asymptotic upper bound on the maximal number of times any
given basic block is executed in the loop.
\end{claim}

These functions may still contain unresolved
functions $g(\vec{p})$ representing loops for which the runtime is not
known statically. We will explain in Section~\ref{sec:hybrid-model} how
we derive these functions empirically from performance measurements. 

\subsection{Compute volume of a full program}

We calculate the compute volume of a full program without recursion.
Any code not part of a loop can be ignored as it has only constant cost.
Similarly, bodies of inner loops can be assumed to have only constant
computational cost since the analysis is inter-procedural and
loop nests are aggregated across function calls.
Therefore, the asymptotic compute volume can be derived by
looking only at the recursively accumulated cost of loop nests.

\begin{theorem}
  Given an application $A$ with a set of input variables V, a set of
  $n$ correctly marked potential performance variables
  $P = (p_1, \dots, p_n) \subseteq V$, no irreducible control flow or
  recursion, the recursive
  accumulation of the iteration volume in each function of the call
  tree (due to no recursion) computes the asymptotic compute volume of
  $A$.
\end{theorem}

The taint analysis therefore yields properties of the
function space of possible performance models but it does not explicitly
generate precise models. In this sense, it provides a ``scaffolding''
that defines some relations among loops. However, the precise function for
each loop is not yet defined. To derive such functions, we first discuss how to
include additional control-flow information into the
model~(\ref{sec:algorithm-selection}). We subsequently refine an empirical
modeling approach to parametrize the missing loop models to derive accurate
overall performance models for each function~(\ref{sec:hybrid-model}).

\subsection{Algorithm selection}
\label{sec:algorithm-selection}

In addition to building a set of performance models for each function discussed
earlier, we apply taint analysis to locate control-flow decisions unrelated
to loop exit conditions that are affected by input parameters.
Instrumenting
conditional branches with taint sinks enables (1) the detection of tainted control-flow
decisions affecting performance models for branches inside any loop
nest and (2) the detection of code paths that
are never visited, including the parameter-based selection of algorithms.

\subsection{Empirical performance modeling is also hard}
\label{sec:hybrid-model}

Our dynamic taint analysis provides us with information on how parameters
influence the compute volume of individual functions, but does not
provide specific functions that describe the asymptotic behavior very
precisely. To close this gap, we build a hybrid analysis by combining
the compute volume information from our compiler-based analysis with a
black-box empirical performance modeler. This modeler runs a program
multiple times with different parameter configurations. Using both
the results of taint analysis and the observed execution times,
it derives a performance-model function that (1) respects parameter
dependencies derived during the taint analysis and (2) provides the best fit
to empirical data. As a black-box performance prediction approach we
use the performance modeling
tool Extra-P~\cite{calotoiu_ea:2013:modeling,shudler_ea:2015,calotoiu_ea:2016}.

\paragraph{Performance function}

A key concept of the Extra-P approach is the {\em performance model normal form} (PMNF), defined in Equation~\ref{eq:epmnf}. It models the effect of parameters $x_{i}$ on a variable of interest $f(x_{1},\dots,x_{m})$, typically execution time or a performance counter.
The PMNF is based on the assumption that performance, at least at
  the level of functions calls, can usually be expressed as a
  combination of polynomial and logarithmic terms. This flexibility in
  expressing behaviors is sufficient to cover most cases encountered
  in practice while keeping the modeling process fast enough to be viable.
%
\begin{equation}
  f(x_{1},\dots,x_{m}) = \sum_{k=1}^n c_k  \cdot  \prod_{l=1}^m x_l^{{i_k}_l} \cdot log_2^{{j_k}_l}(x_l) 
\label{eq:epmnf}
\end{equation}
The PMNF defines a function search space, which is
traversed to find the function that comes closest to
representing the set of measurements. This assumes that the true
function is within the search space.
A possible assignment of all $i_k$ and $j_k$ in a PMNF expression is
called a {\em model hypothesis}. The sets $I, J \subset \mathbb{Q}$
from which the exponents $i_k$ and $j_k$ are chosen and the number of
terms $n$ define the model search space.
The coefficients of all hypotheses are automatically derived
using regression and the hypothesis with the smallest error is chosen to
find the most likely model function. In this work, we use the configuration suggested by Ritter et al.~\cite{ritter_ea:2020:ipdps}.

This approach \emph{always} generates a human-readable expression
  out of any given measurement data. It attempts to explain this data
  as well as possible by fitting the PMNF to the data. The more
  complex the PMNF is, such as by adding more terms or a wider range
  of exponents to the terms, the more freedom the modeling has to fit
  the data. This allows more behaviors to be expressed but risks
  overfitting the data---especially in the presence of noise.

\iftr
\paragraph{Search space}
In practice, a simple selection like the following is often sufficient to
model even complex scientific applications:
\begin{align*}
  n &= 2\\
  I &=\left\{\frac{0}{4},\frac{1}{4},\frac{1}{3},\frac{2}{4},\frac{2}{3},\frac{3}{4},\frac{4}{4},\frac{5}{4},\frac{4}{3},\frac{6}{4},\frac{5}{3},\frac{7}{4},\frac{8}{4},\frac{9}{4},\frac{10}{4},\frac{8}{3},\frac{11}{4},\frac{12}{4}\right\}\\
  J &= \{0, 1, 2\}
\end{align*}
It is possible to expand or modify the sets or
the number of terms if clear expectations regarding the application
behavior exist, but such prior knowledge is not required in the common
case.
For the above process to yield good results, the true function that
is being modeled should not be qualitatively different from what the
normal form can express.  Discontinuities, piece-wise defined
functions, and other behaviors that cannot be modeled by the normal
form will lead to sub-optimal results.
Looking at Equation~\ref{eq:epmnf}, the combinatorial explosion of the
search space for model hypotheses becomes apparent when using the same
assumptions regarding hypotheses generation as we did in the
single-parameter case. With as few as three parameters, the model
search space contains more than $10^{14}$ candidates, making the
search for the best fit a daunting task. Extra-P includes two heuristics that
deal with the excessive size of the search space. The first heuristic speeds
up multi-parameter modeling, as it reduces the search space to only
combinations of the best single-parameter models.  The second
heuristics speeds up model selection for single parameter models.
Combined, the two heuristics allow a search space of hundreds of
billions of models to be reduced to under a thousand~\cite{calotoiu_ea:2016}.
\fi

\paragraph{Limitations}

A significant limitation of Extra-P is the black-box nature of the
approach that uses only empirical measurements to generate performance
models. This means that the models can be affected both by
random noise, and by systemic interference such as network congestion
caused by multiple applications sharing a physical system. While these
effects can be mitigated by repeating measurements and
trying to control the measurement infrastructure, they cannot be eliminated and their impact is
larger the more parameters are
considered~\cite{ritter_ea:2020:ipdps}. In most applications, runtime
is concentrated in a small number of routines, and while these
routines are correctly modeled, the previously discussed disturbances
disproportionately affect regions of code with short runtimes, and in some cases translate to Extra-P effectively modeling noise. Given the large number of such occurrences, in some the noise can randomly resemble a strong correlation between a parameter and a metric. Such false positives can, at the moment, only be eliminated by manual inspection of the code and cost users valuable time.

\paragraph{Hybrid modeler}

Our goal is to allow the PMNF to be as expressive as possible to
   accurately model different performance behaviors
  but wish to prevent this expressivity from generating false positives
  by overfitting. We therefore use taint analysis to define a prior
  for the modeling process in Extra-P.
We use the results of the taint analysis to minimize the negative effects of
measurement noise. The model of computational volume is applied to restrict the
search space by removing parameters that could not affect 
performance.
As a result, the black-box regression algorithm no longer uses non-existing parameter
dependencies in models.

The immediate effect is pruning out parametric models for constant
functions. These functions are notoriously hard to model since the variability
of measurement data forces the modeler to favor functions that are not
constant. The final model is overfitted and likely misleading.
The second important result is the removal of false dependencies in performance
models.
We therefore automate the process of verifying empirical models by removing parameters not present in the code from the search space.

\section{Implementation of \toolname}
\label{sec:perf_taint}

\begin{figure}
  \includegraphics[width=\linewidth]{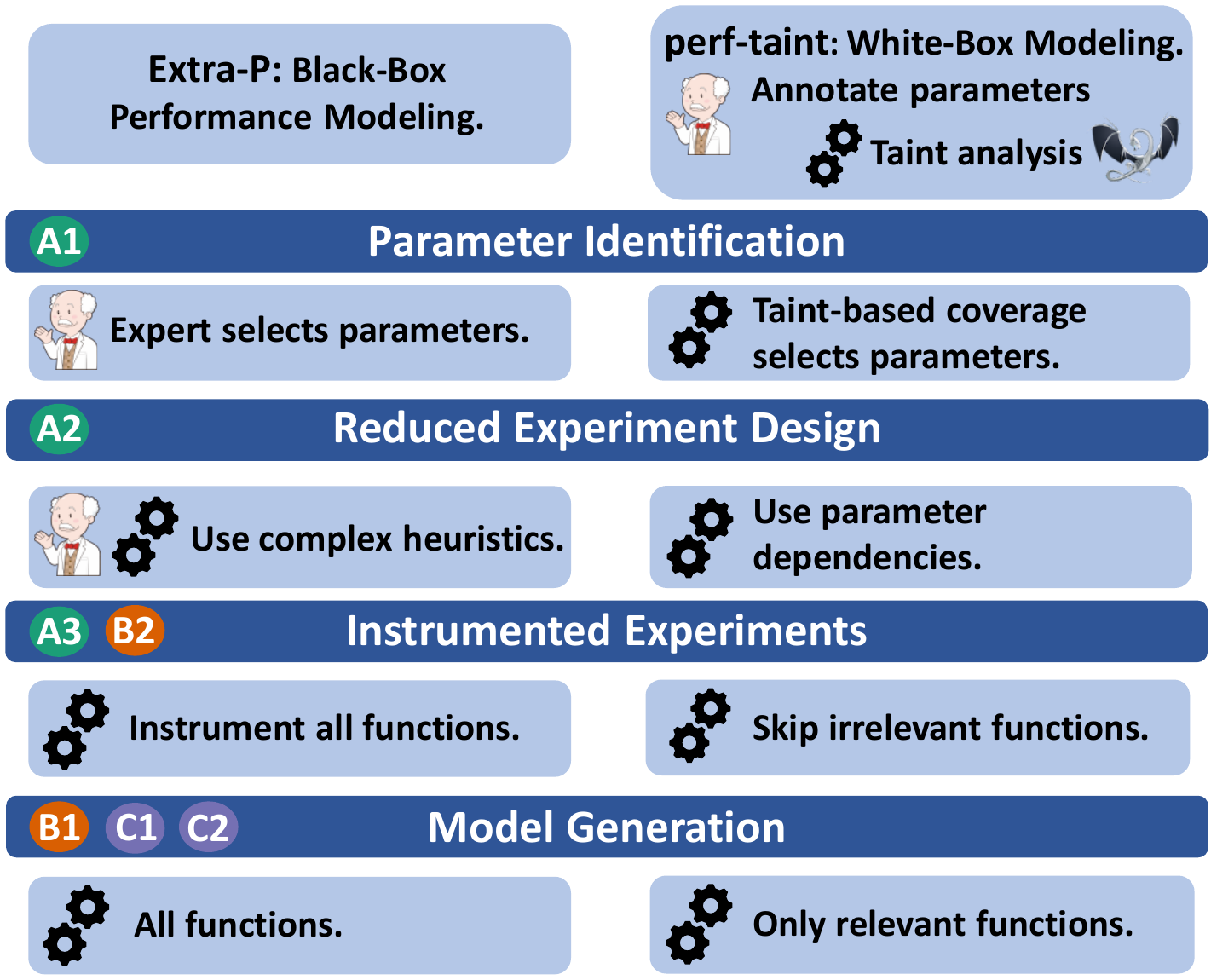}
  \caption{The processing pipeline of \toolname. All four major steps of empirical
  modeling are improved with the program information provided by taint analysis.}
  \label{fig:system-workflow}
\end{figure}

We provide with \toolname an implementation of our performance tainting
  approach, as shown in \autoref{fig:system-workflow}. Our processing pipeline
  includes the new step of tainted modeling that consists of three
  stages: (1) a static
analysis, (2) a dynamic taint analysis; and (3) a database of
performance-critical libraries, which we discuss in detail in the following subsections.
\toolname uses LLVM~\cite{Lattner:2004:LCF:977395.977673} and works on
the level of
the intermediate representation (IR), which makes \toolname
applicable to a range of languages, including C++, Fortran, Julia, etc.
  However, our taint-based modeling approach (\autoref{sec:performance-modeling})
is independent of the taint implementation and can also be built using
other taint-analysis frameworks~\cite{DBLP:conf/ndss/KangMPS11,10.1145/2151024.2151042,she2019neutaint,10.1145/1273463.1273490}.

\autoref{fig:system-workflow} shows how performance
modeling with Extra-P is improved with the program information provided via
taint analysis. Without taint analysis expert knowledge is necessary to decide which parameters have the largest impact on performance and
scalability, a difficult manual task. \toolname
leverages taint analysis to determine how many loops and functions are affected by
each specific parameter, providing a simple yet intuitive coverage metric and removing
from the analysis any parameters that have no effect on performance.
The only user action is the annotation of each input
parameter with one line of code in the program source, as illustrated in the example below.
In contrast to many performance modeling tools~\cite{hoisie-palm-2012,8048922,Lee:2015:CFA:2751205.2751220}, 
we do not require our users to annotate
regions of interests, functions, loop boundaries, or even to provide manually annotated
performance models for each kernel.
\begin{lstlisting}[basicstyle=\scriptsize]
struct cmdLineOpts opts;
ParseCommandLineOptions(argc, argv, myRank, &opts);
register_variable(&opts.nx, "size");
\end{lstlisting}

The next step determines the set of measurements used
for empirical modeling. The user needs to provide the constraints on parameter values,
which is a problem-specific part. The na\"ive approach considers all combinations
of parameter values and therefore the number of samples scales exponentially with the number of parameters.
While the original approach required sophisticated heuristics and a potential accuracy loss to reduce this number,
taint analysis decides which parameter have multiplicative dependencies and which
lead to additive effects. This means often not all combinations are required, therefore reducing the burden of the most computationally expensive part of the pipeline without sacrificing accuracy.
Finally, \toolname uses taint-based information on parameter dependencies to
select only relevant functions for instrumentation and prune models with false
dependencies, leading to better and cheaper models.

\subsection{Static analysis}
\label{sec:static_analysis}

At compile time, we identify all functions that contain no loops or only
loops with constant and statically resolvable trip counts since their performance models
are known to be independent from any program parameter.
To that end, we query an existing loop induction analysis (e.g., ScalarEvolution~\cite{scev}).
%
During this process, we include functions containing library calls that
are known to be affected by performance parameters, such as MPI communication
routines.
 
\subsection{Dynamic taint analysis}
\label{sec:dynamic_analysis}

\ifcnf
\else
We build our solution on top of the DataFlowSanitizer plugin~\cite{dfsan} in the Clang and LLVM compiler toolkit,
a data-flow taint system consisting of (1) a transformation pass applied to source code,
and (2) a runtime library implementing label allocation, query, and union. 
The taint labels are designed as a tree-like data structure, where each label
can represent a union of up to two labels. Each label has an associated 16-bit identifier.
The union operation verifies whether the operands do not represent an equivalent combination
of labels and creates a new one if necessary. While the implementation is less
efficient than a simple bitset solution, it supports up to $2^{16}$ unique labels.
The runtime manages the shadow memory pool that stores taint labels for each memory location of the program. 
The transformation pass instruments each instruction with 
propagation of taint labels from its operands to the output.
\fi

\toolname gathers information on the effects of tainted parameters on each
non-constant loop in the program, by treating loop exit branch conditions
as taint sinks. We store call-path information to distinguish
between function calls that result in different dependencies, letting the empirical
modeler Extra-P create calling-context-aware performance models. As part of post-processing
after program execution, we parse loop nests with parameter dependencies and 
use this information to generate explicit multiplicative and additive dependencies for a function.
The only source of over-approximation in our analysis is the presence of multiple labels
in a single exit condition,
where we conservatively report a multiplicative dependency.
The latter requires more experiments to be accurately parametrized.

\iftr
\paragraph{Control-flow tainting} We extended DataFlowSanitizer with instrumentation for explicit control-flow
tainting since it is necessary
to capture all dependencies in real-world applications. Both the
runtime and the transformation
pass have been extended to store the taint labels associated with control-flow
decisions, and propagate them to variables whose values depend on the
control flow.
Below, we present an example from the LULESH benchmark discussed in~\autoref{sec:case-studies}. The final value
contained in \code{regElemSize} has a control dependence on parameter \code{size} through \code{numElem}
in the loop condition, since the value depends on the number of loop iterations.
This kind of dependency cannot be recovered by only considering data flow, i.e. the propagation of labels
from instruction inputs to an output.
\begin{lstlisting}[basicstyle=\scriptsize]
for(Index_t i = 0 ; i < numElem(); ++i) {
  int r = this->regNumList(i) - 1;
  regElemSize(r)++;
}
\end{lstlisting}

\fi

\subsection{Global state libraries}
\label{sec:global_state_libraries}
Loop-based kernels are not the only way how parameters can affect the performance.
The model has to include parametric effects of communication and synchronization routines.
The performance parameter could affect
their behavior in the following ways: (1) a value tainted by the parameter is exchanged
between processes operating in disjoint memory spaces, (2) the parameter is passed
to the routine explicitly, (3) the parameter is hidden from the user in the library runtime.
We solve issues (2) and (3) by introducing a library database describing
performance-relevant functions, implicit parameters provided by
libraries, and sources of taint values.

We demonstrate the solution on MPI, the most widely-used library for distributed
and high-performance applications. We declare the implicit
parameter \texttt{p}, which denotes the size of the global communicator, and
we include the function \texttt{MPI\_Comm\_size} as a source
of tainted values, writing a label to the memory address passed as a second
argument to the function. We derive parametric dependencies for MPI communication
and synchronization routines from precise analytical models~\cite{Thakur:2005:OCC:2747766.2747771,hoefler-moor-collectives},
and provide them in the library database supplied with \toolname.
As an example, we consider the case of MPI peer--to--peer communication routines.
When they appear in a function, our analysis introduces implicit dependence on \texttt{p}.
Since performance of these functions depends on the network conditions and message size,
we query the taint labels associated with \texttt{count} argument provided to the function and add them
as additional parametric dependencies for this function call.

Taint labels could be transfered between processes withing MPI messages.
The problem of tainting network communication
has already been tackled~\cite{10.1007/978-3-642-25141-2_8},
and an analogous solution for MPI would only have to cover standard MPI datatypes
and the few, well-defined routines that create user-defined datatypes. We
have found that the lack of support for data exchange across
the network is not an obstacle for the applications we analyzed.

\section{Taint Analysis in Action}
\label{sec:case-studies}

\addtolength{\tabcolsep}{-3pt}
\begin{table}\centering
  \small
  \begin{tabular*}{\linewidth}{l @{\extracolsep{\fill}} cc}\toprule
    & \textbf{Piz Daint} & \textbf{Skylake Cluster}\\ \midrule
    CPU & Intel Xeon E5-2695 v4 2.10GHz & Intel Xeon 6154 3GHz\\
    Cores & 2 sockets, 18 cores each & 36 cores\\
    Memory & 128 GB & 384 GB\\
    GCC, MPI & GCC 8.3, Cray MPICH 7.7.2& GCC 8.3, OpenMPI 4.0.3\\
    Software & \multicolumn{2}{c}{Score-P 6.0~\cite{an_mey_ea:2010:cihpc}, Extra-P 3.0~\cite{extra-p}, LLVM 9.0~\cite{Lattner:2004:LCF:977395.977673}}\\
    \bottomrule
  \end{tabular*}
  \caption{Software and hardware configuration of \toolname evaluation on Piz Daint and a local Skylake cluster. }
  \label{tab:systems}
\end{table}
\addtolength{\tabcolsep}{3pt}

\addtolength{\tabcolsep}{-5pt}
\begin{table}\centering
  \small
  \begin{tabular*}{\linewidth}{l @{\extracolsep{\fill}} cc}\toprule
    & \textbf{LULESH} & \textbf{MILC}\\ \midrule
    \textbf{Functions} & 356 & 629 \\
     Pruned Statically/Dynamically & 296/11 & 364/188 \\
     Kernels/Comm. Routines/MPI & 40/2/7& 56/13/8\\
    \textbf{Loops} & 275 & 874 \\
     Pruned Statically & 52 & 96\\
     Relevant & 78 & 196 \\
    \textbf{Modeling}\\
    \textit{p} & $3^{n}(27, ..., 729)$ & $2^{n}(4, ..., 64)$\\
    \textit{size} & 25,30,35,40,45& 32,64,128,256,512\\
    \bottomrule
  \end{tabular*}
  \caption{
  Overview of LULESH and MILC: the two-phase identification of computational kernels, communication routines and MPI functions used, and the manually specified parameter values for two-parameter modeling.
}
  \label{tab:benchmarks}
\end{table}
\addtolength{\tabcolsep}{5pt}

We present the three major categories of improvements that our taint-supported framework brings
to the empirical performance modeling process: decreased cost~(\autoref{sec:cost}),
improved quality~(\autoref{sec:models_quality}),
and the discovery of software and hardware phenomena the knowledge of
which can help validate experiment design and modeling results
\ifcnf
\else
\newline~(\autoref{sec:models_validity}).
\fi
We support our claims by applying taint-supported performance modeling
to two representative HPC benchmarks:
LULESH~\cite{LULESH2:changes} and MILC~\cite{Bernard:1991:SQG:2748634.2748643},
summarized in~\autoref{tab:benchmarks} and discussed in the next two
paragraphs. The hardware and software systems used are summarized
in~\autoref{tab:systems}.

\paragraph{LULESH}

is a scientific application written in C++, implementing
stencil computations for a hydrodynamic shock problem on a
three-dimensional mesh. The code is structured around the main class
\code{Domain} and contains multiple simple methods. Their expected
constant computational effort is hard to capture empirically because
the presence of noise makes timing data unreliable for such short
functions.
We run the taint analysis of this application with \code{size} 5 and 8 MPI ranks,
leaving other parameters at the default value, since it provides 
a representative execution of the application that is close to 
parameter configurations used in modeling~(\autoref{sec:models_quality}).
If we choose the number of MPI ranks \code{p} and the grid size \code{size} as
performance-model parameters, a typical use case, our analysis marks
86.2\% of the functions as not influenced by these two parameters,
allowing the immediate classification of their models as constant.


\paragraph{MILC}

We model the performance of the su3\_rmd application from the MIMD Lattice Computation,
a collection of scientific applications working on problems from the lattice quantum
chromodynamics (QCD) fields.
%
We analyze here the effects of two parameters frequently chosen
for scaling studies: (1) the \code{size} of
the space-time domain, which is computed from the four parameters
\code{nx, ny, nz}, and \code{nt}, and (2) \code{p}, the number of
MPI ranks.
We apply the taint analysis of this application
with a \code{size} of 128 on 32 MPI ranks. Again, the taint analysis
identifies 87.7\% of the functions as constant relative to these two
parameters.
This corrects 77\% models previously
  indicating performance effects. Our analysis is
  confirmed by the preceding manual analysis and the validation with up to
  512,000 processes~\cite{milc-modeling}.

\addtolength{\tabcolsep}{-4pt}
\begin{table}\centering
  \small
  \begin{adjustbox}{width=\columnwidth,center}
  \begin{tabular*}{\linewidth}{l @{\extracolsep{\fill}} c ccccccc}\toprule
    \textbf{LULESH\ }  & Total & \textit{p} & \textit{size} & \textit{regions} & \textit{iters} & \textit{balance} & \textit{cost} & \textbf{\textit{p, size}} \\ 
    \midrule
    Functions & 43 & 2 & 40 & 13 & 4 & 9 & 2 & 40\\
    Loops & 86 & 2 & 78 & 27 & 4 & 20 & 2 & 78\\
    \midrule
    \multirow{2}{*}{\textbf{MILC}} & \multirow{2}{*}{Total} & \multirow{2}{*}{\textit{p}} & \multirow{2}{*}{\textit{size}} & \multirow{2}{*}{\textit{trajecs}} & \textit{warms} & \textit{nrest.} & \textit{mass},\textit{beta} & \multirow{2}{*}{\textbf{\textit{p, size}}} \\
    & & & & & \textit{steps} & \textit{niter} & \textit{nfl.} / \textit{u0} & \\
    \midrule
    Functions & 56 & 54 & 53 & 12 & 9 & 6 & 1 / 4& 56\\
    Loops & 196 & 187& 161 & 39 & 31 & 15 & 1 / 7 & 196\\
  \end{tabular*}
  \end{adjustbox}
  \caption{Computational kernels and loops in multi-parameter modeling.
      \textbf{\textit{p, size}} is not equal to the sum of corresponding columns since multiple parameters can affect the same region.
  }
  \label{tab:params}
\end{table}
\addtolength{\tabcolsep}{4pt}

\let\oldsubsection\thesubsection
\let\oldsubsubsection\thesubsubsection
\renewcommand\thesubsection{\caseicon{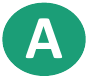}}
\titleformat{\subsection}{\normalfont\normalsize\itshape}{\thesubsection}{1em}{}
\subsection[6.A Cost]{Cost}
\label{sec:cost}
\renewcommand\thesubsection{\oldsubsection}
The cost of the modeling process is influenced by two major factors:
the number of the required performance experiments, which significantly
grows with the number of model parameters, and the cost of these
experiments under instrumentation. Tainted performance modeling lowers
these costs in multiple ways while reducing the dependence on human
expertise. First, it supports automatic pruning of the parameter
space~(\autoref{sec:parameter_pruning}), avoiding many unnecessary
experiments. It can expose parameter dependencies at an
early stage~(\autoref{sec:parameter_dependencies}), allowing smarter
experiment design with even less experiments. Finally, we show that
the ability to judge the performance relevance of a function upfront
\ifcnf
\else
substantially decreases the instrumentation
overhead~(\autoref{sec:instrumentation_overhead}).
\fi

\renewcommand\thesubsubsection{\caseicon{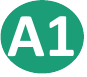}}
\titleformat{\subsubsection}{\normalfont\normalsize\itshape}{\thesubsubsection}{1em}{}
\subsubsection{Parameter pruning}
\label{sec:parameter_pruning}

\begin{figure}[H]
  \centering
  \includegraphics[width=\linewidth]{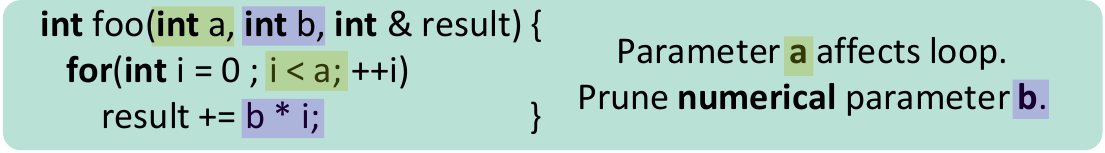}
\end{figure}
High-performance computing applications often involve a large set of
execution parameters. In practice, 
limited compute budgets restrict the number of model parameters to
three, and even with boundless resources one should not go much
beyond, as the impact of noise would become too strong~\cite{rl_model}.
Tainting allows us to decide which parameter influences which part of
the program. Programmers should mark program parameters found in routines
parsing command-line arguments and configuration files.
Our analysis determines all parameters without effect on the
control flow and counts the number of loops and functions directly
affected by a specific parameter.~\autoref{tab:params} summarize parameter
pruning on both benchmarks, excluding communication routines relevant only because of calls to MPI.

LULESH includes six major parameters: the problem size \code{size},
the number of MPI ranks \code{p}, \code{regions},
\code{balance}, \code{cost}, \code{iters}. To build a
two-parameter model providing the broadest coverage of performance relevant functions our taint analysis suggests we select \code{size} and \code{p}.

In MILC, we detect the performance-relevant
parameters \code{nx}, \code{ny}, \code{nz}, \code{nt}, \code{steps}, \code{niter},
\code{warms}, \code{trajecs} and the implicit parameter
\code{p}. Our findings are identical with the ground truth established
by experts in a laborious manual
process~\cite{milc-modeling}.

\begin{figure*}[htb]
	\centering
  \includegraphics[width=\linewidth]{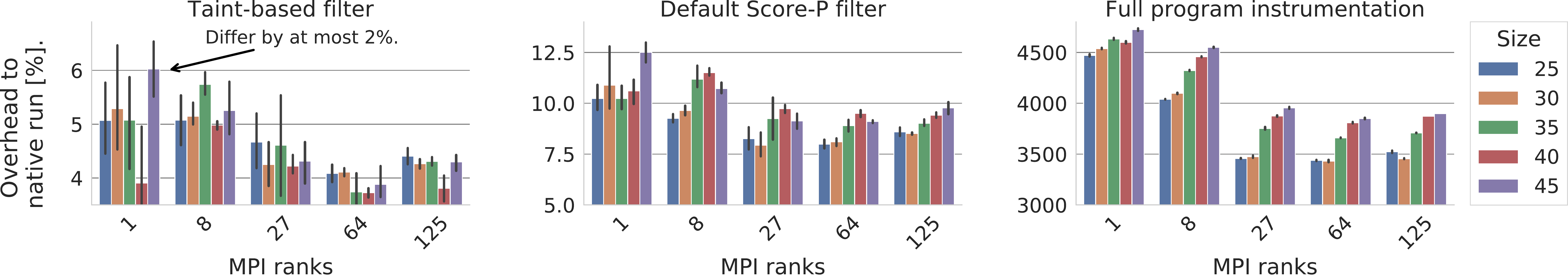}
  \caption{The Score-P instrumentation overhead of LULESH on the Skylake cluster. Plot scales
            are tuned to enhance visibility.}
  \label{fig:overhead_lulesh}
\end{figure*}
\iftr
\begin{figure*}[htb]
	\centering
  \includegraphics[width=\linewidth]{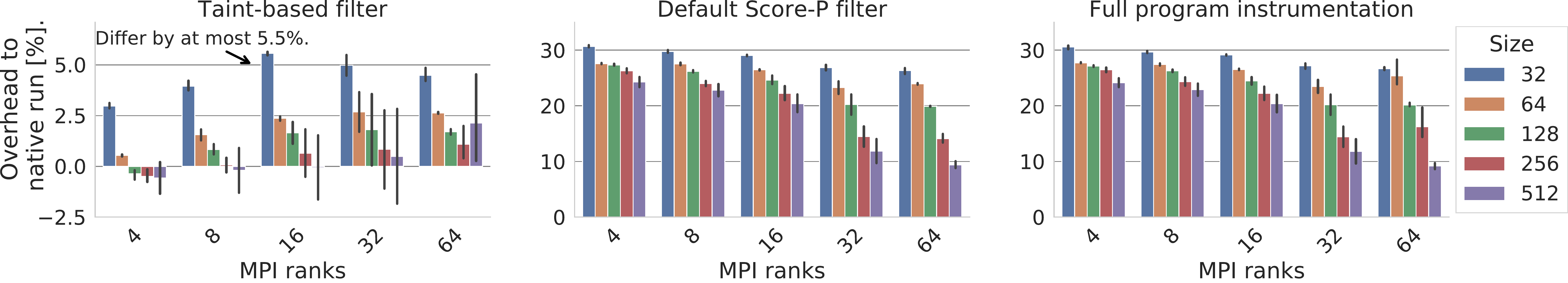}
  \caption{The Score-P instrumentation overhead of MILC on the Skylake cluster. Plot scales
            are tuned to enhance visibility.}
  \label{fig:overhead_milc}
\end{figure*}
\fi

\renewcommand\thesubsubsection{\caseicon{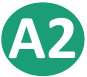}}
\titleformat{\subsubsection}{\normalfont\normalsize\itshape}{\thesubsubsection}{1em}{}

\subsubsection{Parameter dependencies}
\label{sec:parameter_dependencies}

\begin{figure}[H]
  \centering
  \includegraphics[width=\linewidth]{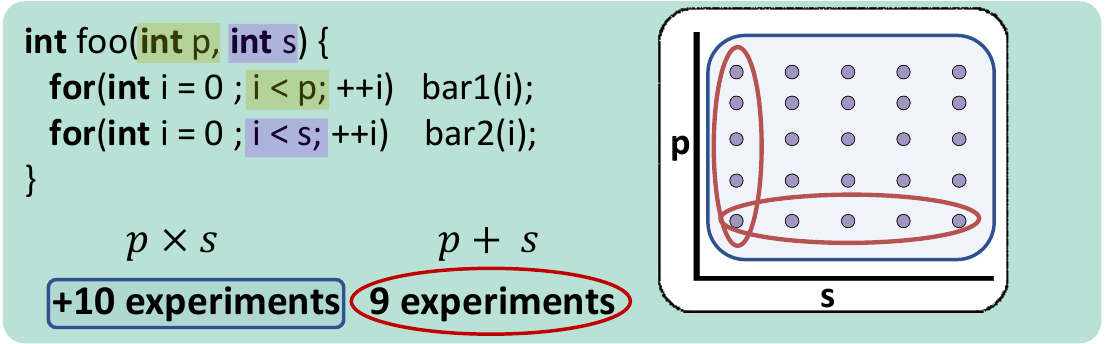}
\end{figure}
Taint analysis can find parameter dependencies, such as
multiplicative dependencies between parameters influencing the
iteration count in outer and inner loops,
and additive dependencies between parameters influencing the iteration
count of non-nested loops.
For routines where parameter dependencies are detected as additive only, accurate performance models can be generated by creating single parameter models for each of the parameters involved. Should this be true for all routines in an applications, the experiment design as a whole can be simplified and its dimensionality reduced.

An interesting corner case of a multiplicative dependency is LULESH,
where the taint-based modeling detects a single instance of the
parameter \code{iters} in the main loop of the program. Through that
we recover a multiplicative dependency with all other model
parameters. The number of iterations therefore linearly affects the
entire computation. We can reduce the dimensionality of sample space,
since \code{iters} does not grant useful insights into application performance.
	
\renewcommand\thesubsubsection{\caseicon{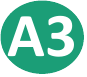}}
\titleformat{\subsubsection}{\normalfont\normalsize\itshape}{\thesubsubsection}{1em}{}
\subsubsection{Instrumentation overhead}
\label{sec:instrumentation_overhead}

\begin{figure}[H]
  \centering
  \includegraphics[width=\linewidth]{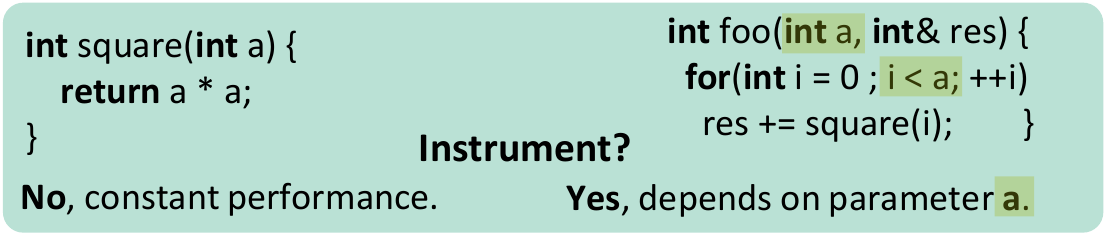}
\end{figure}
In the default instrumentation mode, Score-P~\cite{an_mey_ea:2010:cihpc}, a widely used measurement infrastructure and the default for Extra-P, estimates
whether a function should be inlined and therefore excluded from
instrumentation.
This approach is inappropriate for empirical performance modeling
because it might encourage the compiler to remove performance-critical
functions through inlining, obscuring potential sources of bottlenecks and
impeding effective performance analysis. Thus, without contextual
information from the taint analysis, each function must be conservatively
assumed to be influenced by changing parameter values, leading to
instrumentation of all functions and significant runtime overhead.

Using the results of our analysis, we decrease the instrumentation
overhead by instrumenting selectively, including only functions
that are affected by a parameter change. In particular, we 
prune most of the simple constant functions, such as class getters and
setters, which are irrelevant to scaling studies, without reducing the
model quality for the remaining functions. We compare the overhead of Score-P
with default, full, and our selective instrumentation.

The results for LULESH in Figure~\ref{fig:overhead_lulesh}
clearly demonstrate how severe the overhead can become for C++ applications. Depending on the number of ranks and the problem size, removing constant and irrelevant functions \emph{decreases the execution time by a factor of up to 45 times}. Although the overhead of the default Score-P instrumentation is manageable, the results may influence the models themselves. Our selective instrumentation contains 40 important application functions, 
while the default Score-P run instruments less than half of the
performance-relevant functions but includes helper functions with constant runtime. 
\ifcnf
\else
The results for MILC are presented in Figure~\ref{fig:overhead_milc}.
There, the geometric mean of overheads are $1.6\%$ for selective
instrumentation and $23\%$ for full and default instrumentation.
The default instrumentation provides little to no benefit, whereas
the overheads of our instrumentation are negligible on larger-scale runs.
\fi
We observe that selective instrumentation provides the most significant
improvements in C++ applications, which are gaining popularity in the HPC community.

Since the default Score-P instrumentation misses important functions, causing
false-negative results, the modeling process can only use a full instrumentation mode.
The core-hour costs of taint analysis are 1 and 16 hours for LULESH and MILC, respectively,
while the costs of the experiment decreased from 20483 to 547 hours for LULESH (97.3\%),
and from 364 to 321 hours for MILC (13.4\%), when switching from a \textit{full} to \textit{taint-based} instrumentation.
The savings from reduced overhead significantly outweigh the costs of an additional analysis.
	
%

\renewcommand\thesubsection{\caseicon{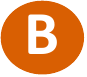}}
\titleformat{\subsection}{\normalfont\normalsize\itshape}{\thesubsection}{1em}{}

\subsection[6.B Quality]{Quality}
\label{sec:models_quality}
We use tainting to mitigate the effects of measurement noise\\
~(\autoref{sec:model_noise}), and the selective instrumentation discussed in~\autoref{sec:instrumentation_overhead} to reduce
the intrusion of instrumentation, improving the quality of the resulting models~(\autoref{sec:model_intrusion}).
	
\renewcommand\thesubsubsection{\caseicon{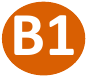}}
\titleformat{\subsubsection}{\normalfont\normalsize\itshape}{\thesubsubsection}{1em}{}
\subsubsection{Noise resilience}
\label{sec:model_noise}

\begin{figure}[H]
  \centering
  \includegraphics[width=\linewidth]{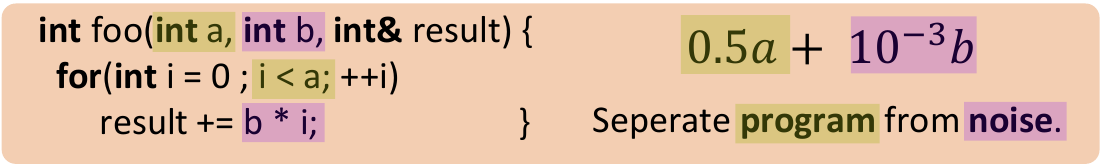}
\end{figure}
We apply the model obtained by the taint analysis
to the model estimation in Extra-P, to prune
models with false dependencies and evaluate the validity of experiments.
We combine the five values of each parameter defined in~\autoref{tab:benchmarks} to
construct a set of training data with 25 points, repeating each
measurement five times to reduce the effects of random noise,
resulting in 125 measurements, which we obtain by using up to 21 (LULESH) and 2 (MILC) Piz Daint nodes.

We compare the new models to black-box ones. We generally observe that models generated using taint analysis are
closer to (nearly always exactly matching) the ground truth that
we established with manual performance modeling techniques
using code inspection~\cite{Hoefler:2011:PMS:2063348.2063356}.
We select for the
comparison only those functions whose data sets do not contain
values with a coefficient of variance larger than $0.1$, as they are too affected by noise to be reliable.


We compare our findings for MILC with models created
manually~\cite{milc-modeling} as a ground truth. For the kernels
manually studied, the taint analysis correctly identifies the
dependencies on both \code{p} and \code{size} in accordance with the
theoretical study. The empirical model
also converges to the same model for each function.
There are four
\code{MPI\_Comm\_Rank} functions which we correctly detect as constant
where measurement noise previously caused incorrect models to be
generated.

\renewcommand\thesubsubsection{\caseicon{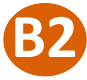}}
\titleformat{\subsubsection}{\normalfont\normalsize\itshape}{\thesubsubsection}{1em}{}
\subsubsection{Less intrusion}
\label{sec:model_intrusion}

\begin{figure}[H]
  \centering
  \includegraphics[width=\linewidth]{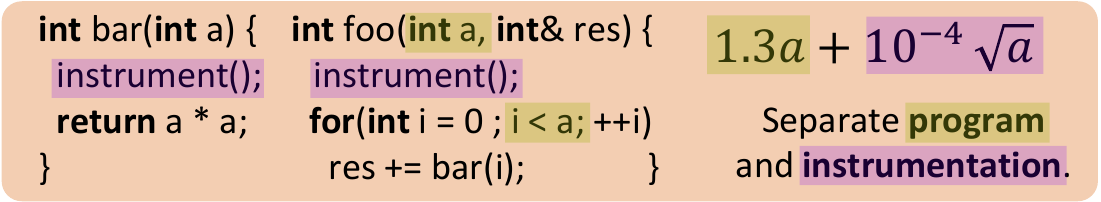}
\end{figure}
Empirical performance modeling relies on measurements. The instrumentation
process introduces overhead, increasing the cost of the experiments, as discussed in the previous section.
Yet, even more troubling is that the instrumentation perturbs the
measurements, causing the resulting models to change qualitatively. We
compare models from the fully instrumented code with those from code
where only the routines identified as performance relevant are
instrumented. Beyond the observation that nearly all runtimes are
almost two orders of magnitude bigger under full instrumentation,
critical routines such as \code{CalcQForElems} show different models
depending on the degree of instrumentation. The model derived from
fully instrumented runs shows an additive dependency between $p$ and $size$, $3\cdot 10^{-3} \cdot p^{0.5}+ 10^{-5}\cdot {size}^3$ while the filtered instrumentation shows a multiplicative dependency $ 2.4\cdot 10^{-8}\cdot p^{0.25}\cdot {size^3}$. The second model is validated by previously determined models~\cite{calotoiu_codesign:2018}, providing a strong argument for using a targeted approach towards instrumentation rather than simply instrumenting full applications. The default Score-P filter does not instrument this function, leading to false-negative result in this case.

\renewcommand\thesubsection{\caseicon{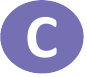}}
\titleformat{\subsection}{\normalfont\normalsize\itshape}{\thesubsection}{1em}{}
\subsection[6.C Validity]{Validity}
\label{sec:models_validity}

The empirical approach we study always generates a performance model
from a given input. We previously discussed in
Section~\ref{sec:models_quality} how we can make sure that we generate
the best possible model. There are situations, however, where the
systemic influence of hardware or poor experiment design make the data
unsuitable for understanding algorithmic performance. We identify such cases and 
provide guidance to identify the cause of the issue. We discuss two such examples: the effect of hardware
contention in a multi-core system~(\autoref{sec:hardware_effects}),
and a qualitative change in the modeled function across the
experiment~(\autoref{sec:model_design}).

\renewcommand\thesubsubsection{\caseicon{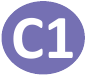}}
\titleformat{\subsubsection}{\normalfont\normalsize\itshape}{\thesubsubsection}{1em}{}
\subsubsection{Detecting hardware contention}
\label{sec:hardware_effects}

\begin{figure}[H]
  \centering
  \includegraphics[width=\linewidth]{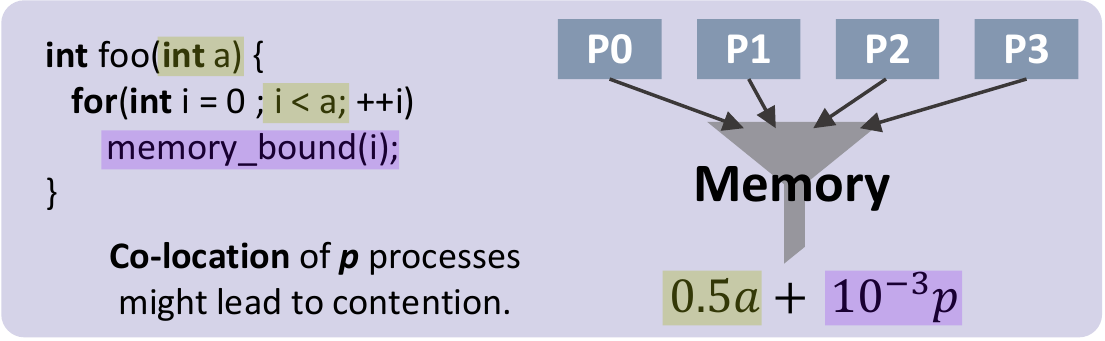}
\end{figure}
We evaluated the black-box and white-box modeling approaches with measurements of LULESH from~\autoref{tab:benchmarks}. We observed a significant number of
computational kernels, where the new model is worse at fitting the
data even though it no longer contains false dependencies on
\code{p}. Since the taint analysis proved that such functions cannot
include such a dependency, yet it is visible in measurements, we
conclude that that the resulting performance model must be affected by factors outside the application code itself. The taint-based modeling pipeline \textbf{detects} the presence of this perturbation while it was unknown to the black-box modeling approach. 

\begin{figure}
	\centering
  \includegraphics[width=\linewidth]{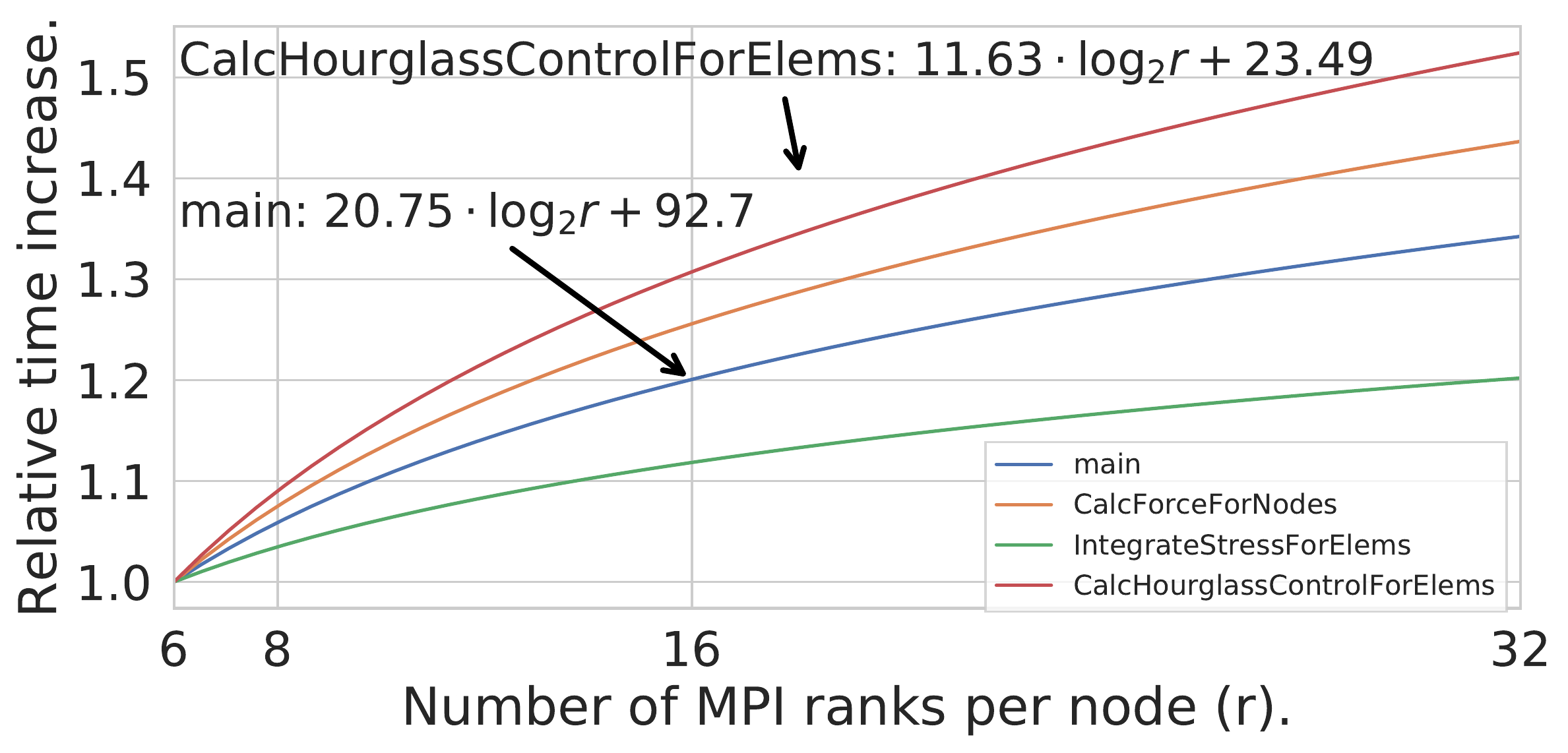}
  \caption{The one-parameter model of LULESH evaluating the effect of the number of MPI ranks \code{r} per node. Computational kernels experience slowdown because of hardware contention when many processes occupy the same socket.}
  \label{fig:contention-plot}
\end{figure}

We formulate the hypothesis that the co-location of MPI ranks on the same socket leads to hardware contention effects on functions with no dependence on MPI ranks in the source code. We test this hypothesis with a new experiment keeping the number of MPI ranks and problem size constant ($\code{p}=64$ and $\code{size}=30$) and varying the numbers or MPI ranks per node
\code{r}, scaled from 2 to 18. By disabling multithreading, the larger number of cores available to each MPI rank should not affect the performance of compute kernels, and only communication routines might benefit from optimized MPI operations when processes are co-located. The expectation is that non-communication routines should have constant models. The entire application shows a significant increase in execution time, by 50\%, from $130 \si{s}$ to $195 \si{s}$, with the corresponding model $2.86 \cdot \log_2^2{\code{r}} \si{s} + 127 \si{s}$. Out of 73 functions, 31 have an increasing model with statistically sound measurements. \autoref{fig:contention-plot} shows a few major examples.

Given the significant number of memory operations in the program, the
saturation of memory bandwidth is the most likely culprit. Thanks to
the inclusion of program information from taint analysis, we provide a
type of insight that has not been available with purely black-box performance modeling. Modeling results that are independent of hardware effects and parallel allocations are possible for LULESH but only for certain levels of node saturation with MPI processes.

\renewcommand\thesubsubsection{\caseicon{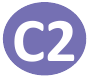}}
\titleformat{\subsubsection}{\normalfont\normalsize\itshape}{\thesubsubsection}{1em}{}
\subsubsection{Validating the experiment design}
\label{sec:model_design}

\begin{figure}[H]
  \centering
  \includegraphics[width=\linewidth]{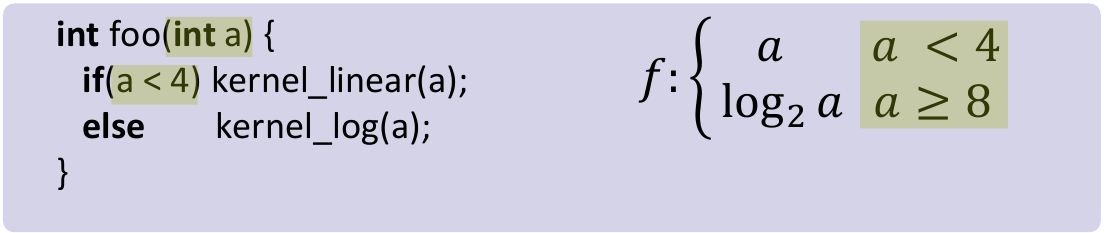}
\end{figure}
We evaluated the black-box and white-box modeling approaches with MILC test runs
and noticed the largest difference between models of communication routines
such as \code{MPI\_Isend} and a MILC internal implementation of the gather collective operation.

Although the measurements
are statistically valid, they fail to present a consistent behavior across the
modeling domain. We notice a \textit{qualitative}, not merely a quantitative
difference between execution on 4, 8, 16 and larger numbers of ranks. As there is
more than one behavior to be modeled in one interval, the parametric models estimated
by Extra-P cannot represent the function accurately unless more measurement data is provided~\cite{ilyas_ea:europar:2017}. 

We have expanded our taint analysis to provide information regarding branches of code that are executed or not executed and therefore where application and/or library behavior can qualitatively change. This empowers the user to appropriately design his experiments to ensure there is only one behavior present in the data.

\renewcommand\thesubsection{\oldsubsection}
\renewcommand\thesubsubsection{\oldsubsubsection}

\section{Conclusion}
\label{sec:conclusion}
This work is the first to show that taint analysis, a method
originally introduced to track the flow of sensitive information in
computer programs, can be used to significantly advance the state of
the art in empirical performance modeling for HPC applications. We
showed various use-cases to improve cost, quality, and validity of the
resulting models. Cost is reduced by lowering the number of necessary
experiments as well as making individual experiments
cheaper. Moreover, higher noise resilience and less
instrumentation-induced intrusion render the models more
accurate. Finally, with its ability to approximate the computational
volume of program executions, taint analysis can help expose
contention effects that prolong the runtime beyond what one would
expect from considering computational volumes alone. Overall, our
results show that for applications of realistic complexity empirical
modeling must be carefully combined with static and compiler-assisted
dynamic analyses to deliver high-quality and actionable performance
models.

\begin{acks}
  This work has been funded by the Deutsche Forschungsgemeinschaft (DFG, German Research Foundation) 
  through Grants WO 1589/8-1 and WO 1589/10-1,
  and by the Schweizerische Nationalfonds zur Förderung der wissenschaftlichen Forschung
  (SNF, Swiss National Science Foundation) through
  Project 170415, programme Ambizione
  (Grant PZ00P2168016) and programme Spark (Grant CRSK-2\_190359/1),
  We would also like to thank the Swiss National Supercomputing Centre (CSCS)
  for providing us with access to their supercomputing machines Daint and Ault.
\end{acks}


\bibliographystyle{ACM-Reference-Format}
\bibliography{paper}

\end{document}